\documentstyle[12pt]{article}
\begin{document}
\large
\begin{center}
{\bf About virtual $\pi \leftrightarrow K$ Meson Oscillations}\\
\vspace{2cm}
Kh.M. Beshtoev\\
\vspace{2cm}
Joint Institute for Nuclear Research, Joliot Curie 6\\
141980 Dubna, Moscow region, Russia\\
\vspace{2cm}
\end{center}
\par
{\bf Abstract}\\

\par
In the framework of the Standard Model the probability (and time) of
$\pi \leftrightarrow K$ transitions (oscillations) are computed.
These transitions are virtual ones since masses of $\pi$ and $K$ mesons
differ considerably.
These transitions (oscillations) can be registered through
$K$ decays after transitions of virtual $K$ mesons
to their own mass shell by using their
quasielastic strong interactions. But for avoiding the background from
inelastic $K$ mesons, the energies $E_\pi$ of $\pi$ mesons
must be less than the
threshold energy of their creation, i.e. $E_\pi < 0.91 $ GeV. The optimal
distances for observation of these oscillations are computed.
Solution of the problem of origin of mixing angle in the
theory of vacuum oscillation is given.

\section{Introduction}

\par
The vacuum oscillation of neutral $K$ mesons is well investigated at
the present
time [1]. This oscillation is the result of $d, s$ quark mixings and
is described by Cabibbo-Kobayashi-Maskawa matrices [2]. The angle
mixing $\theta$ of neutral $K$ mesons is $\theta = 45^O$ since $K^o,
\bar K^o$
masses are equal (see $CPT$ theorem). Besides, since their masses are equal
these oscillations are real ones, i.e. their transitions to each other go
without suppression. The case when  two particles with different
masses having widths oscillate
was discussed in works [3]. Now it is necessary to discuss
$\pi \leftrightarrow K$ oscillations as an oscillation example of particles
having large  difference of masses.
The $\pi \leftrightarrow K$ transitions
(or oscillations) arise from the  existence of quark mixings
($d' = cos \theta d + sin \theta s$).
The same transitions (or oscillations) arise in the model of
dynamical analogy of Cabibbo-Kobayashi-Maskawa matrices [4].
\par
Let us pass to detailed discussion of the $\pi \leftrightarrow K$ transitions.

\section {The $\pi \stackrel{W sin\theta}\leftrightarrow K$ Transitions
(Oscillations)}

\par
2.1 {\bf $\pi \stackrel{W} \leftrightarrow \pi$ Transitions}
\par
We begin the discussion of the problem of $\pi \stackrel{W} \leftrightarrow
K$ transitions (or oscillations) with consideration of transitions
(oscillations)
of $\pi \stackrel{W} \leftrightarrow \pi$ through $W$ bosons,
i.e. through the following Feinmann diagram:\\

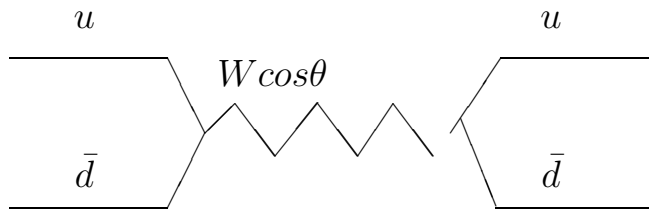
\unitlength=1.00mm
\special{em:linewidth 0.4pt}
\linethickness{0.4pt}
\begin{picture}(113.00,50.00)
\put(29.00,39.00){\line(1,0){20.00}}
\put(49.00,39.00){\line(1,-2){5.00}}
\put(54.00,29.00){\line(-1,-2){5.00}}
\put(49.00,19.00){\line(-1,0){20.00}}
\put(29.00,19.00){\line(0,0){0.00}}
\put(29.00,19.00){\line(0,0){0.00}}
\put(29.00,19.00){\line(0,0){0.00}}
\put(29.00,19.00){\line(0,0){0.00}}
\put(29.00,19.00){\line(0,1){0.00}}
\put(54.00,29.00){\line(1,1){4.00}}
\put(58.00,33.00){\line(3,-4){5.33}}
\put(63.33,26.00){\line(4,5){5.67}}
\put(69.00,33.00){\line(3,-4){5.33}}
\put(74.33,26.00){\line(2,3){4.67}}
\put(79.00,33.00){\line(3,-4){5.33}}
\put(84.33,26.00){\line(5,6){2.33}}
\put(86.67,29.00){\line(2,3){6.67}}
\put(93.33,39.00){\line(1,0){19.67}}
\put(88.00,31.00){\line(2,-5){4.67}}
\put(92.67,19.00){\line(1,0){20.33}}
\put(28.00,39.00){\line(1,0){1.00}}
\put(28.00,19.00){\line(1,0){2.00}}
\put(38.00,44.00){\makebox(0,0)[cc]{$u$}}
\put(38.00,24.00){\makebox(0,0)[cc]{$\bar d$}}
\put(63.00,37.00){\makebox(0,0)[cc]{$W cos\theta$}}
\put(100.00,44.00){\makebox(0,0)[cc]{$u$}}
\put(100.00,24.00){\makebox(0,0)[cc]{$\bar d$}}
\end{picture}

\par
The amplitude $M$ of this process (after using the standard procedure [5])
has the form
$$
M = \frac{G}{\sqrt{2}} cos \theta F_\alpha Q^\alpha =
\frac{g^2_W}{8 m^2_W} F_\alpha Q^\alpha  ,
\eqno(1)
$$
where $F_\alpha = f_\pi \phi_\pi p_\alpha$, $Q^\alpha = \bar d_L
\gamma^\alpha u_L$ , $\phi_\pi$ is $\pi$-meson wave function and $\mid
\phi_\pi \mid_2 = 1 $, $f_\pi$ is $\pi$ constant decay ($f_\pi \cong 125$
GeV), $p_\alpha$ is $\pi$ four-momentum, $Q^\alpha$ is quark current,
$\theta$ is Cabibbo's quark mixing angle.
\par
If we use the expression $\hat p_\pi = \hat p_u + \hat p_{\bar d}$ and
the Dirac equation
$$
(\hat p - m) u = 0
$$
(where $u$ is quark wave function), then one can rewrite Eq.(1) in the
following form:
$$
M = \frac{G}{\sqrt{2}} cos \theta f_\pi \phi_\pi (m_u + m_{\bar d}) \bar d_L
\gamma_5 u_L .
\eqno(2)
$$
Using the standard procedure for $\bar {{\mid M \mid}^2}$, one obtains
the following expression:
$$
\bar {\mid M \mid^2} = G^2 cos^2 \theta f^2_\pi (m_u + m_{\bar d})^2
4(p_{\bar d} p_u) \cong
$$
$$
\cong 4 G^2 cos^2 \theta f^2_\pi (m_u + m_{\bar d})^2 m_\pi E_u ,
\eqno(3)
$$
where $E_u \cong E_d \cong \frac{m_\pi}{2}$.
\par
Then the transition $\pi \stackrel{W} \rightarrow \pi$ probability
$W(...)$ is
$$
W(\pi \stackrel{W}\rightarrow \pi) = \frac{\bar {\mid M \mid^2}}{2 m_\pi}
\int \frac{d^3 p_u}{2 E_u (2 \pi)^3} \frac{d^3 p_u}{2 E_d (2 \pi)^3}
(2 \pi)^4 \delta( p_u + p_{\bar d} - p_\pi) =
$$
$$
= \frac{\bar {\mid M \mid^2}}{4 \pi m_\pi}
\int \delta(E_{\bar d} + E_u - m_\pi)
\frac{E_{\bar d} dE_{\bar d}}{E_u} \cong
$$
$$
\cong \frac{\bar {\mid M \mid^2}}{4 \pi} \frac{E}{m^2_\pi} ,
\eqno(4)
$$
where $E \cong \frac{m_\pi}{2}$  .
\par
Then using the expression for $G^2_W$, one can rewrite
equation (4) in the form
$$
W(\pi \stackrel{W}\rightarrow \pi) \cong
\frac{G^2 f^2_\pi cos^2 \theta (m_u + m_{\bar d})^2 m_\pi}{8\pi}  =
\eqno(5)
$$
$$
= \left( \frac{g_W^2}{4 \sqrt{2} m_W^2} \right)^2
\frac{f^2_\pi cos^2 \theta (m_u + m_{\bar d})^2 m_\pi}{8\pi} =
$$
$$
= W_0(\pi \rightarrow \mu \nu_\mu) (\frac{m_u + m_{\bar d}}{m_\mu})^2 ,
$$
where
$$
W_0(\pi \rightarrow \mu \nu_\mu) =
\frac{G^2 f^2_\pi cos^2 \theta (m_\mu)^2 m_\pi}{8\pi}
$$
and
$$
\tau'_0 = \frac{1}{W_0(... )} .
$$
\par
Then the time $\tau(...)$ of $\pi \stackrel{W} \rightarrow \pi$
transition is
$$
\tau(\pi \stackrel{W}\rightarrow \pi) =
\frac{1}{W(\pi \stackrel{W}\rightarrow \pi)} .
\eqno(6)
$$
\\
2.2 {\bf $\pi \stackrel{W sin \theta} \rightarrow K$ Transitions
(Oscillations)}\\
\par
The diagram for $\pi \stackrel{W sin \theta} \longrightarrow K$ transitions
when one takes into account $d, s$ quark mixings and $W$ exchange has
the form \\

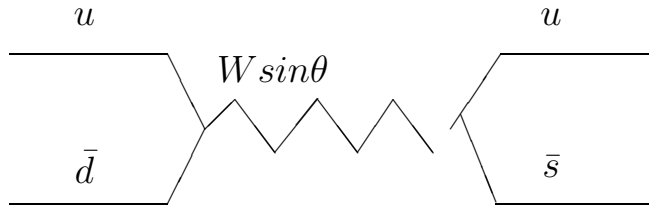
\unitlength=1.00mm
\special{em:linewidth 0.4pt}
\linethickness{0.4pt}
\begin{picture}(113.00,50.00)
\put(29.00,39.00){\line(1,0){20.00}}
\put(49.00,39.00){\line(1,-2){5.00}}
\put(54.00,29.00){\line(-1,-2){5.00}}
\put(49.00,19.00){\line(-1,0){20.00}}
\put(29.00,19.00){\line(0,0){0.00}}
\put(29.00,19.00){\line(0,0){0.00}}
\put(29.00,19.00){\line(0,0){0.00}}
\put(29.00,19.00){\line(0,0){0.00}}
\put(54.00,29.00){\line(1,1){4.00}}
\put(58.00,33.00){\line(3,-4){5.33}}
\put(63.33,26.00){\line(4,5){5.67}}
\put(69.00,33.00){\line(3,-4){5.33}}
\put(74.33,26.00){\line(2,3){4.67}}
\put(79.00,33.00){\line(3,-4){5.33}}
\put(84.33,26.00){\line(5,6){2.33}}
\put(86.67,29.00){\line(2,3){6.67}}
\put(93.33,39.00){\line(1,0){19.67}}
\put(88.00,31.00){\line(2,-5){4.67}}
\put(92.67,19.00){\line(1,0){20.33}}
\put(28.00,39.00){\line(1,0){1.00}}
\put(28.00,19.00){\line(1,0){2.00}}
\put(38.00,44.00){\makebox(0,0)[cc]{$u$}}
\put(38.00,24.00){\makebox(0,0)[cc]{$\bar d$}}
\put(63.00,37.00){\makebox(0,0)[cc]{$W sin\theta$}}
\put(100.00,44.00){\makebox(0,0)[cc]{$u$}}
\put(100.00,24.00){\makebox(0,0)[cc]{$\bar s$}}
\end{picture}

\par
It is clear that at $d, s$ mixings the transition from
the mass shell of $\pi$ meson does not take place  ,
i.e. $K$ meson created from $\pi$ meson
remains on the mass shell of $\pi$ meson.
\par
Repeating the same calculations in (2)--(5) for
$\pi \stackrel{W sin\theta} \leftrightarrow K$ transitions
for the probability $W(...)$ of $\pi \stackrel{W sin\theta} \rightarrow K$
transitions, one obtains the following expression:
$$
W(\pi \stackrel{W sin\theta} \rightarrow K) \cong
\frac{G^2 f^2_\pi sin^2 \theta (m_u + m_{\bar d})^2 m_\pi}{8\pi}  =
\eqno(7)
$$
$$
= \left(\frac{g^2_W}{4\sqrt{2} m^2_W} \right)^2
\frac{f^2_\pi cos^2 \theta (m_u + m_{\bar d})^2 m_\pi}{8\pi} .
$$
Then the time $t$ of $\pi \stackrel{W sin\theta} \rightarrow K$
transition is
$$
\tau(\pi \stackrel{W sin\theta}\longrightarrow K) =
\frac{1}{W(\pi \stackrel{W sin\theta}\longrightarrow K)} .
\eqno(8)
$$
\par
The relation between the time $\tau(\pi, \pi)$
of $\pi \stackrel{W cos\theta} \rightarrow \pi$ transition and the time
$\tau(\pi, K)$ of $\pi \stackrel{W sin\theta} \rightarrow K$ transition is
$$
\frac{\tau(\pi \stackrel{W sin\theta}\rightarrow K)}
{\tau(\pi \stackrel{W cos\theta}\rightarrow \pi)} = \frac{1}{tg^2\theta}  .
\eqno(9)
$$
\par
So, at the $\pi \leftrightarrow K$ transitions, $\bar s$ remains on the mass
shell of $\bar d$ quark (i.e. $K$ is on $\pi$ mass shell) and then $K$ mesons
transit back into $\pi$ mesons and this process goes on the background of
$\pi$ decays. It is clear that these oscillations
(transitions) are virtual ones and can be seen through $K$
meson decays if virtual $K$ mesons transit to their own mass shell.
Since $K$
mesons take part in the strong interactions, one can do it through
their quasiinelastic strong interactions.
This problem will be considered
in the next work.
\par
Now we pass to computation of the probability  of $\pi \leftrightarrow
K$ oscillations.

\section{Probability of $\pi \stackrel{W sin\theta}\longrightarrow K$
(Virtual) Oscillations}
\par
The mass matrix of $\pi$ and $K$ mesons has the form
$$
\left(\begin{array}{cc} m_\pi & 0 \\ 0 & m_K  \end{array} \right) .
\eqno(10)
$$
\par
Due to the presence of strangeness violation in the weak interactions,
a nondiagonal term appears in this matrix
and then this mass matrix is transformed in
the following nondiagonal matrix:
$$
\left(\begin{array}{cc}m_\pi & m_{\pi K} \\ m_{\pi K} & m_K  \end{array}
\right)   ,
\eqno(11)
$$
which is diagonalized by turning through the angle $\beta$ and
$$
tg 2\beta = \frac{2m_{\pi K}}{\mid m_\pi - m_K \mid}   ,
$$
$$
sin 2\beta = \frac{2m_{\pi K}}{\sqrt{(m_\pi - m_K)^2 +(2m_{\pi K})^2}}  .
\eqno(12)
$$
\par
It is interesting to remark that expression (12) can be obtained from the
Breit-Wigner distribution [3]
$$
P \sim \frac{\Gamma/2)^2}{(E - E_0)^2 + (\Gamma/2)^2}
\eqno(13)
$$
by using the following substitutions:
$$
E = m_K,\hspace{0.2cm} E_0 = m_\pi,\hspace{0.2cm} \Gamma/2 = 2m_{\pi K} ,
\eqno(14)
$$
where $\Gamma \equiv W(... )$.
\par
Now we can consider two cases of $\pi, K$ oscillations-real and virtual:
\par
1. If we consider the real transition of $\pi$ into $K$ mesons then
$$
sin^2 2\beta \cong \frac{4m^2_{\pi K}}{(m_\pi - m_K)^2} \cong 0 ,
\eqno(15)
$$
i.e. the probability of the real transition of $\pi$ mesons into $K$ mesons
through
weak interaction is very small since $m_{\pi K}$ is very small.
\par
How can we understand this real $\pi \rightarrow K$ transition?
\par
If $2m_{\pi K} = \frac{\Gamma}{2}$ is not zero, then it means that the
mean mass of $\pi$ meson is $m_\pi$ and this mass is distributed by $sin^2
2\beta$ (or by the Breit-Wigner formula) and the probability of
the $\pi \rightarrow K$
transition differs from zero.
So, this is  a solution of the problem of
origin of mixing angle in the theory of vacuum oscillation.
\par
2. If we consider the virtual transition of $\pi$ into $K$ meson then, since
$m_\pi = m_K$,
$$
tg 2\beta = \infty  ,
$$
i.e. $\beta = \pi/4$, then
$$
sin^2 2\beta = 1     .
\eqno(16)
$$
\par
We will consider the second case since it is of real interest.
\par
If at $t = 0$ we have the flow $N(\pi, 0)$ of
$\pi$ mesons then at $t \ne 0$ this flow will decrease since $\pi$ mesons
decay and then we have the following flow $N(\pi, t)$ of $\pi$ mesons:
$$
N(\pi, t) = exp(- \frac{t}{\tau_0}) N(\pi, 0)  ,
\eqno(17)
$$
where $\tau_0 = \tau'_0 \frac{E_\pi}{m_\pi}$.
\par
One can express the time $\tau(\pi \leftrightarrow K)$ through the time
of $\tau_0$, then
$$
\tau(\pi \stackrel{W sin \theta} \longrightarrow K) =
\tau_0 (\frac{m_\mu}{m_u + m{\bar d}})^2 \frac{1}{tg^2 \theta} .
\eqno(18)
$$
\par
The expression for the flow $N(\pi \rightarrow K, t)$, i.e. probability of
$\pi$ to $K$
meson transitions at time $t$, has the form
$$
N(\pi \rightarrow K, t) =
N(\pi, t) sin^2 \left[\frac{\pi t}{\tau(\pi \stackrel{W} \longrightarrow K)}
\right ] =
$$
$$
= N(\pi, 0) exp(-\frac{t}{\tau_0})sin^2\left[\frac{\pi t}{\tau_0}
\frac{tg^2 \theta}{(\frac{m_\mu}{m_u + m_{\bar d}})^2}\right]           .
\eqno(19)
$$
\par
Since $\tau(\pi \rightarrow K) >> \tau_0$ at
$t = \tau(\pi \rightarrow K)$ nearly all $\pi$ mesons will decay, therefore
to determine a more effective time (or distance) for observation of
$\pi \rightarrow K$ transitions
it is necessary to find the extremum of
$N(\pi \rightarrow K, t)$, i.e. Eq. (19):
$$
\frac{d N(\pi \rightarrow K, t)}{d t} = 0 .
\eqno(20)
$$
From Eqs. (19) and (20) one obtain the following equation:
$$
\frac{2 \pi tg^2 \theta}{(\frac{m_\mu}{m_u + m_{\bar d}})^2}  =
tg\left[\frac{t \pi tg^2 \theta}{\tau_0 (\frac{m_\mu}{m_u + m_{\bar d}})^2}
\right] .
\eqno(21)
$$
\par
If one takes into account that the argument of the right
part of (21) is a very small value, one can rewrite the right part
of (21) in the form
$$
tg\left[\frac{t \pi tg^2 \theta}{\tau_0 (\frac{m_\mu}{m_u + m_{\bar
d}})^2}\right] \cong \frac{t \pi tg^2 \theta}{\tau_0(\frac{m_\mu}{m_u +
m_{\bar d}})^2} .
\eqno(22)
$$
Using (21) and (22) one obtains that the extremum
of $N(... )$ takes place at
$$
\frac{t}{\tau_0} \cong 2 \hspace{1cm} or
\hspace{1cm}t \cong 2 \tau_0 .
\eqno(23)
$$
And the extremal distance $R$
for observation of $\pi \rightarrow K$ oscillations is
$$
R =  t v_\pi \cong 2 \tau_0 v_\pi  ,
\eqno(24)
$$
and the equation for $N(\pi \rightarrow K,
2\tau_0)$ has the following form:
$$
N(\pi \rightarrow K, 2\tau'_0) = N(\pi, 0)
exp(-2)sin^2\left[2 \pi \frac{tg^2 \theta}{(\frac{m_\mu}{m_u + m_{\bar
d}})^2}\right] \cong
\eqno(25)
$$
$$
\cong N(\pi,0) 5.1\hspace{0.2cm}{10}^{-6} ,
$$
where $m_u + m_{\bar d} \cong 15$ \hspace{0.2cm} MeV,
\hspace{0.2cm} $tg^2 \theta \cong 0.048$.
\par
Let us pass to discussion of kinematics of $K$ meson creation processes.\\

\section{Kinematics Processes of $K$ Meson Creation }
\par
So, if one has $\pi$ mesons, then with the probability determined by Eq.
(19) they virtually transit into $K$ mesons and if these virtual $K$ mesons
participate in quasielastic strong interactions then they become real $K$
mesons. Then through $K$ meson decays one can verify this process.
\par
The energy threshold $E_{thre, \pi}$ of the quasielastic reaction
$\pi^+ + p \rightarrow K^+ + p$ is
$$
E_{thre, \pi} = 0.61 \hspace{0.2cm} GeV.
$$
\par
Besides, this quasielastic reaction of $\pi$ mesons can create $K$ mesons
in inelastic reactions. An example of this inelastic reaction is
the following reaction:
$$
\pi^{+} +  n  \rightarrow K^+ + \Lambda  .
\eqno(26)
$$
The energy threshold $E^{inel}_{thre, \pi}$ is
$$
E^{inel}_{thre, \pi} = 0.91 \hspace{0.2cm}GeV .
$$
\par
To avoid the problem with $K$ mesons created in inelastic reaction,
one must take energies $E_\pi$ of $\pi$ mesons less than $0.91$ GeV,
i.e. $E_\pi$ must be
$$
0.61 \leq E_\pi \leq 0.91 \hspace{0.2cm}GeV .
\eqno(27)
$$
\par
The optimal distances for observation of $\pi \leftrightarrow K$
oscillations can be computed using Eqs.(24) and (27).

\section{Conclusion}

In the framework of the Standard Model the probability
(and time) (see Eq.(19))
of the $\pi \leftrightarrow K$ transitions (oscillations) were computed.
These transitions are virtual ones since masses of $\pi$ and $K$ mesons
differ considerably.
These transitions (oscillations) can be registered through
$K$ decays after transitions of virtual $K$ mesons to their own
mass shell by  using
their quasielastic strong interactions. But for avoiding the background of
inelastic $K$ mesons the energies $E_\pi$ of $\pi$ mesons
must be less than the
threshold energy of their creation, i.e. $E_\pi < 0.91 $ GeV. The optimal
distances for obsevation of these oscillations were computed
(see Eq. (24)).
Solution of the problem of origin of mixing angle in the
theory of vacuum oscillation was given.
\par
The computation of $\pi \leftrightarrow K$ oscillations can also be performed
in the framework of the model of dynamical analogy of
Cabibbo-Kobayashi-Maskawa matrices [4] through $B\pm$ boson exchanges.
At low energetic limit the results (time and length of oscillations) are the
same.

\begin{center}
REFERENCES  \\
\end{center}
\par
\medskip
\noindent
1. Review of Particle Prop., Phys. Rev. 1992, D45, N. 11.
\par
\noindent
2. Cabibbo N., Phys. Rev. Lett., 1963, 10, p.531.
\par
Kobayashi M. and Maskawa K., Prog. Theor. Phys., 1973, 49,
\par
p.652.
\par
\noindent
3. Beshtoev Kh.M., JINR Commun. E2-93-167, Dubna, 1993;
\par
   JINR Commun. E2-95-326, Dubna, 1995;
\par
   Chinese Journ. of Phys., 1996, 34, p.979.
\par
\noindent
4. Beshtoev Kh.M., JINR Commun. E2-94-293, Dubna, 1994;
\par
   Turkish Journ. of Physics, 1996, 20, p.1245;
\par
   JINR Commun. E2-95-535, Dubna, 1995;
\par
   JINR Commun. P2-96-450, Dubna, 1996.
\par
   JINR Commun. E2-97-210, Dubna, 1997.
\par
\noindent
5. Okun L.B., Leptons and Quarks, M., Nauka, 1990.

\end{document}